\documentclass{article}

\usepackage{microtype}
\usepackage{graphicx}
\usepackage{subfigure}
\usepackage{booktabs} 

\usepackage{hyperref}

\usepackage[accepted]{icml2025}

\usepackage{amsmath}
\usepackage{amssymb}
\usepackage{mathtools}
\usepackage{amsthm}

\usepackage[capitalize,noabbrev]{cleveref}

\theoremstyle{plain}

\theoremstyle{definition}

\theoremstyle{remark}

\usepackage[textsize=tiny]{todonotes}

\icmltitlerunning{AI, Brain Death Detection, and Islamic Law}

\begin{document}

\twocolumn[
\icmltitle{AI, Brain Death Detection, and Islamic Law}



\icmlsetsymbol{equal}{*}

\begin{icmlauthorlist}
\icmlauthor{Muhammad Aurangzeb Ahmad}{yyy}
\end{icmlauthorlist}

\icmlaffiliation{yyy}{Department of Computer Science, University of Washington Bothell, Bothell, USA}

\icmlcorrespondingauthor{Muhammad Aurangzeb Ahmad}{maahmad@uw.edu}

\icmlkeywords{brain death, Islamic jurisprudence, disorders of consciousness, bayyina, machine learning, AI ethics}

\vskip 0.3in
]



\printAffiliationsAndNotice{}  

\begin{abstract}
The deployment of machine learning systems capable of detecting covert
consciousness in neurologically injured patients creates a profound challenge at the intersection of clinical medicine, AI ethics, and Islamic
jurisprudence.  We argue that  the
shift from binary clinical verdicts to probabilistic, temporally granular
neural-state estimates should be addressed through three foundational constructs in Islamic
legal epistemology: \textit{bayyina} (clear evidentiary proof),
\textit{yaq\={\i}n} (epistemic certainty), and the theologically mandated
agnosticism about the \textit{r\={u}\d{h}} (soul).  We survey the current technical literature
on AI-based consciousness detection, map it onto the 
landscape of Islamic brain death scholarship, and identifykey challenges. We also discuss its implicationsfor AI surrogate decision systems.
\end{abstract}

\section{Introduction}
\label{sec:intro}
 
In 2006, \citet{owen2006detecting} demonstrated that a patient in a
behaviourally unresponsive vegetative state could follow commands by
wilfully modulating her brain's metabolic activity as measured by
functional MRI. This was a patient who, by every prior clinical criterion, had no
detectable inner life.  This landmark finding opened a field that has since
revealed covert consciousness in up to 15--20\% of patients formally
diagnosed as unresponsive
\citep{monti2010willful,giacino2014disorders,schnakers2009diagnostic}. The domain has since been transformed by machine learning and AI. Deep neural networks
trained on hundreds of thousands of EEG, electrocorticography (ECoG), and
local field potential (LFP) samples now achieve high accuracy in predicting
consciousness levels across diverse neurological populations, with results
correlating significantly with gold-standard behavioural measures in held-out
validation data \citep{ai_consciousness_biorxiv2024,frontiers_doc2025}.
Multimodal approaches, fusing resting-state fMRI functional connectivity,
quantitative EEG features, diffusion tractography, and PET metabolic
imaging through classification models are now being
evaluated in multi-centre European clinical trials \citep{perbrain2024}.
 
This technical progress carries profound and unexamined consequences for ethical issues in end of life medical care, especially around brain death
and organ transplantation.  The AI community has discussed algorithmic
fairness, data bias, and clinical-deployment ethics extensively, but has not
engaged the world's second-largest religious tradition, Islam, on such questions. This paper makes the following contributions:
\begin{enumerate}
  \item We survey the current technical literature on AI-based DoC detection
    for ML readers unfamiliar with its clinical stakes.
  \item We map these developments onto the  Islamic jurisprudential
    landscape on brain death.
  \item We propose a cross-disciplinary research agenda for ML researchers and
    Islamic bioethicists.
\end{enumerate}

A note on scope: this paper addresses Sunni jurisprudential frameworks. Shia maraji positions on brain death differ in important respects, most notably in the greater weight accorded to the heartbeat criterion and in the specific rulings of Grand Ayatollah Sīstani and others within the Twelver tradition. This would require separate treatment.
 
\section{The Technical Landscape: What AI Now Detects}
\label{sec:technical}
 
\subsection{Disorders of Consciousness: Taxonomy and Stakes}
 
Disorders of consciousness (DoC) refer to neurological conditions, typically
resulting from severe acquired brain injury, in which the relationship
between neural activity and behavioural responsiveness is disrupted.  The
principal diagnostic categories are as follows: (i)~\emph{coma}: absence of wakefulness
and awareness; (ii)~\emph{unresponsive wakefulness syndrome} (UWS),
formerly vegetative state: eyes open, sleep-wake cycles preserved, no
behavioural evidence of awareness; (iii)~\emph{minimally conscious state}
(MCS): reproducible but inconsistent behavioural signs of awareness; and
(iv)~\emph{cognitive motor dissociation} (CMD): intact neural
command-following with no behavioural output.
 
Given that we are dealing with a problem where the stake are very high, misdiagnosis has profound consequences for withdrawal-of-treatment
decisions, family communication, pain management, and organ-donation
eligibility. Diagnostic error rates between UWS and MCS
have historically been estimated at approximately 40\% using bedside
behavioural assessment alone \citep{andrews1996misdiagnosis,schnakers2009diagnostic}.
 
\subsection{AI-Powered Detection: State of the Art}
 
A 2025 scoping review of the DoC-ML literature identified 49,417 candidate
articles across PubMed, Embase, Scopus, and Cochrane Library
\citep{frontiers_doc2025}.  High-quality studies span supervised ML (SVM,
XGBoost, random forests), deep learning (CNNs, LSTMs, graph neural networks),
and hybrid IoT-ML approaches for continuous bedside monitoring. The following is a high level overview of some of the apprahces that are used in this domain. It is meant to be non-exhasustive and we only discuss the represenattive example given the limitations of space.
 
\textbf{EEG-based approaches:}
Quantitative EEG (qEEG) features, including microstate analysis, P300 signal
detection, fractal-dimension analysis, and functional-connectivity
metrics, combined with ML classifiers achieve high accuracy in DoC
classification \citep{frontiers_doc2025}.  \citet{digregorio2022} showed that
combining multiple EEG biomarkers with ML enhances predictive accuracy beyond
any single biomarker.
 
\textbf{Deep learning for consciousness detection:}
\citet{ai_consciousness_biorxiv2024} describe a generative-discriminative
architecture in which deep convolutional neural networks (DCNNs), trained on
over 680,000 EEG, ECoG, and LFP samples from humans and animals, are pitted
against biophysically grounded dynamical brain models in an adversarial loop.
The DCNN trained on acute TBI coma patients achieved statistically significant
correlation with GCS scores in held-out validation ($p < 0.0001$), and with
CRS-R scores in chronic DoC patients ($p < 0.0001$).  It should be noted that the system
produces \emph{probabilistic continuous outputs} and not necessarily binary classifications.
 
\textbf{Multimodal fusion:}
The EU-funded PerBrain Consortium is conducting multi-centre trials combining
EEG, resting-state fMRI (rs-fMRI), diffusion MRI, anatomical MRI, and
FDG-PET, analysed through classification models across three European
sites \citep{perbrain2024}.  \citet{yang2024rsfmri} used rs-fMRI to measure
inter-regional connectivity, identifying subtle neural patterns indicative of
awareness in patients with DoC.
 
\textbf{Brain-muscle network analysis:}
\citet{flo2025} demonstrated that combining EEG with electromyography (EMG)
and cardiac recordings through a network-analysis approach detects covert
command-following, with heart activity and cortical power jointly predicting
mental movement rehearsal. This is a promising direction for ICU bedside deployment.
 
\textbf{TMS-EEG perturbational complexity.}
Transcranial magnetic stimulation combined with EEG can assess the capacity
for consciousness via the perturbational complexity index (PCI) without
requiring task performance. This provides a measurement independent of sensory
processing or motor output \citep{casali2013}.
 
\subsection{Covert Consciousness}

An important finding in recent years is the detection of
\emph{cognitive motor dissociation} (CMD). This refers to covert cortical command-following
in patients whose bedside behavioural assessment indicates no awareness. Task-based fMRI and EEG can reveal preserved covert consciousness in up to 14–25\% of UWS patients  \citep{degruyter2024} \citep{Bodien2024CMD}.
 
A positive result is strong evidence of awareness. However, a negative result cannot
rule out awareness\citep{claassen2024,young2024}.  Researchers therefore proposed
the term ``covert cortical processing'' rather than ``covert consciousness''
for patients with intact cortical responses to passive stimuli but no
discernible active-task responses. This framing acknowledges an epistemic gap between
neural measurement and subjective experience that no technology can close
\citep{pmc_covert2024}.
 
The gap that exists here is the permanent underdetermination of subjective experience by
neural measurement.  This
is a fundamental feature of the hard problem of consciousness
\citep{chalmers1995} that neither deeper networks nor richer imaging
modalities will dissolve.  We revisit this issue in Section~\ref{sec:mismatches}.
 
\section{The Jurisprudential Landscape: Brain Death in Islamic Law}
\label{sec:jurisprudence}
 
\subsection{Classical Definition of Death}
 
Classical Islamic jurisprudence defined death through observable cessation of
breath (\textit{nafas}) and heartbeat. These are the signs that are accessible to the senses,
requiring no instrumentation.  The \textit{r\={u}\d{h}} (spirit/soul)
understood as the animating principle breathed by God into the human being
(\textit{wa nafakhtu f\={\i}hi min r\={u}\d{h}\={\i}}: ``and I breathed into
him of My spirit,'' Qur\textsuperscript{'}an 15:29), departs at death and
returns to God. Crucially, the Qur\textsuperscript{'}an explicitly forecloses human knowledge
of the \textit{r\={u}\d{h}}'s nature: ``They ask you about the spirit. Say:
the spirit is from the command of my Lord, and you have been given of
knowledge only a little'' (Qur\textsuperscript{'}an 17:85).  For Muslims, this
also meant a type of agnosticism about the nature of the soul and a theological instruction
and also the limits of human reason The eleventh-century scholar al-Ghaz\={a}l\={\i} formulated the canonical
position in Islamic thought: separation of the soul from the body is the end of its dominance
over the body, and this event, not cardiac or respiratory function per se,
constitutes death \citep{albar2012}.  Some Muslim scholars even argued that the soul's relationship to its substrate is therefore theologically prior to, and cannot be reduced to, any biological criterion, however sophisticated. Death in classical \textit{fiqh} is thus \emph{defined} by empirical signs
while its \emph{meaning}, the departure of the
\textit{r\={u}\d{h}}, transcends empirical determination.  This structural
duality, latent in every pre-modern discussion of death, is important when we discuss the use of instrumentation in determination of life and death.
 
\subsection{The Brain Death Controversy}
 
When the Harvard Ad Hoc Committee formalised neurological criteria for death
in 1968 \citep{harvard1968}, Islamic scholars faced a new  problem to address.  The new criteria described a state in which the heart continued to
beat, the body remained warm, and the organism appeared superficially alive;
yet apnoea, absent brainstem reflexes, and (in some protocols) electrocerebral
silence were taken to indicate the irreversible loss of the person.
As a response to this challenge, three positions have emerged.
 
\textbf{Position~1: Brain death is true death:}
The irreversible destruction of the brain's integrative capacity constitutes
the loss of the human person as a whole.  \citet{albar2012} and
\citet{champasha2017} argue that the absence of \textit{nafs} (personhood)
and \textit{nafas} (breath) in apnoeic coma constitutes departure of the
\textit{r\={u}\d{h}}, and that the ventilated body is maintained artificially
rather than alive in a meaningful sense.  The Islamic Organisation of Medical
Sciences (IOMS, 1986) and the Islamic Fiqh Academy of the Organisation of
Islamic Cooperation (OIC, 1987) issued resolutions broadly aligned with this
view, enabling transplantation \textit{fatw\={a}s} across Gulf states and
much of the Arab world.  Notably, Saudi Arabia's High Committee on Brain Death
requires EEG confirmation before establishing the diagnosis. This is an acknowledgement that the standard clinical examination is insufficient without additional
neurophysiological corroboration \citep{albar2012}.
 
\textbf{Position~2: Brain death is not death:}
The heartbeat is the sign of life; a perfused body is a living body.  Death
occurs when the heart irreversibly stops.  \citet{rady2018} argue that the
secular concept of neurological death was constructed specifically to enable
organ procurement and represents a novel imposition on Islamic communities
that is at odds with the Qur\textsuperscript{'}anic definition of death as
biological disintegration.  On this view, withdrawing ventilation from a
brain-dead patient constitutes an act of killing, and many South Asian Deobandi
and Barelvi scholars, as well as several Shia \textit{mar\={a}ji\textsuperscript{c}}
including Grand Ayatollah S\={\i}st\={a}n\={\i}, align with this position
\citep{padela2015}.  Jurists in this camp further dispute whether the brain is
the ``seat'' of the soul at all, a prerequisite for the whole-brain
formulation of death \citep{padela2015}. A full analysis of Shia positions, including the institutional \textit{fatwa} landscape is beyond the scope of this paper.
 
\textbf{Position~3: Deliberate agnosticism:}
A third position holds that the theological injunction to humility about the
\textit{r\={u}\d{h}} (Qur\textsuperscript{'}an 17:85) demands that Islamic
jurisprudence not issue a binding ruling.  Uncertainty is itself theologically
appropriate; the diversity of \textit{fatw\={a}} positions should be preserved
as principled epistemic humility rather than resolved by scholarly majority
vote.  This view underpins the refusal of some \textit{fuqah\={a}'} to issue
any ruling on organ procurement from brain-dead patients, not because they
reject it but because they hold the evidentiary threshold for such a
determination to be inherently unattainable.
 
\textbf{Current landscape.}
The problem of proof of consciousness and death has implications for things like organ donation. A systematic review of  \textit{fatw\={a}s} around this domain found that organ donation
is broadly permitted within Islamic law, though conditions and scope vary
significantly across schools and regions \citep{champasha2017}.  The Fiqh
Council of North America (FCNA) issued a detailed ruling in 2018 concluding that brain death satisfies
Islamic criteria provided specific clinical safeguards are met \citep{fcna2020}.
The Islamic juridical deliberations around brain death largely took place over
twenty-five years ago; the debates within Muslim bioethics require both
updating and deepening with regard to these early rulings \citep{padela2015}.
 
\subsection{The Epistemological Framework}
\label{sec:epistemo}
 
Islamic legal epistemology provides a graded taxonomy of certainty that
directly governs what kinds of evidence can support what kinds of legal
conclusions.  The relevant grades are: \textit{yaq\={\i}n} (certainty,
knowledge admitting no doubt); \textit{\d{z}ann gh\={a}lib} (dominant
probability, sufficient for most legal purposes under normal circumstances);
\textit{shakk} (doubt, which suspends or blocks action); and \textit{wahm}
(mere conjecture, which is legally ineffective). It should be noted that given the complexity and depth of Islamic jurisprudence, it is not possible to cover this topic given the limited space. We however give a high level overview of some of the concepts in Islamic \textit{fiqh} relevant to our current discussion.
 
\textit{Bayyina}  iterally ``that which makes clear'', is the evidentiary
standard for facts carrying grave legal consequences, including death
certification, criminal conviction, and the dissolution of marriage.
Classical \textit{fiqh} construed \textit{bayyina} narrowly as personal
testimony by qualified witnesses; contemporary scholars have debated whether
the concept can accommodate scientific evidence not based on human testimony,
as discussed in Section~\ref{sec:precedents}.  The principle of
\textit{i\d{h}tiy\={a}\d{t}} (precaution) requires that where a doubt
concerning a life exists and cannot be resolved, the more protective
conclusion must prevail.  This operates as a structural constraint. Thus, t is not
sufficient to show that brain death is probably present, one must show that
the possibility of remaining life has been positively excluded.
 
Two further maxims are relevant.  \textit{Al-yaq\={\i}n l\={a} yaz\={u}lu
bil-shakk} (certainty is not removed by doubt) means that the established
presumption of life persists until overturned by evidence of equivalent
certainty. \textit{\d{D}ar\={u}ra}
(necessity) can relax evidentiary standards under acute exigency, and has
been invoked to justify organ procurement but \textit{\d{d}ar\={u}ra}
doctrines are bounded (\textit{tuqaddar bi-qadrih\={a}} necessity is
assessed proportionately), and cannot be extended to routinise what began
as an exceptional permission.
 
The question this paper raises is: \emph{what evidentiary grade does
an AI probabilistic consciousness score achieve?}  As we argue in
Section~\ref{sec:mismatches}, the answer is that it achieves neither
\textit{yaq\={\i}n} nor \textit{bayyina}. However,  it may do more than merely
generate \textit{shakk}, because it provides a quantified, actionable estimate
of residual awareness that the precautionary framework must take into account.
 
\subsection{Precedents: When Islamic Law Has Engaged\\New Diagnostic Technology}
\label{sec:precedents}
 
The problem of new diagnostic technology disrupting established evidentiary
categories is not novel to Islamic jurisprudence.  Three precedents illuminate
the structural challenge AI monitoring now poses, and suggest both the
resources available to Islamic law and the limits of those resources.
 
\textbf{Blood transfusion and the \textit{dar\={u}ra} model:}
Blood transfusion presented one of the first major intersections of modern
medical technology and Islamic law.  The Qur\textsuperscript{'}an prohibits
the consumption of blood (Qur\textsuperscript{'}an 5:3). Scholars thus debated
whether transfusion, blood introduced into the body by a different
route, fell under this prohibition.  Beginning in 1959, the Grand Mufti of
Egypt and the Grand Mufti of Tunisia both issued \textit{fatw\={a}s}
permitting transfusion under \textit{\d{d}ar\={u}ra}: necessity permits what
is otherwise prohibited when life is at stake \citep{ncbi_bioethics2015}.
This set the template for Islamic bioethical engagement with new medical
technology: (i) identify the relevant Qur\textsuperscript{'}anic or
\textit{Sunna} prohibition; (ii) assess whether the new technology falls
within its scope; (iii) if so, ask whether \textit{\d{d}ar\={u}ra} applies.
The blood transfusion case resolved relatively cleanly because the question
was binary i.e., permitted or not, and the life-saving benefit was direct and
unconditional.  AI consciousness monitoring does not resolve cleanly on this
model: the technology does not save lives by its use. It may, under Islamic
precautionary logic, \emph{prevent} deaths by blocking premature
certification, but this protective function conflicts with the
\textit{ma\d{s}la\d{h}a} of organ availability.
 
\textbf{DNA evidence and the reconstruction of \textit{bayyina}:}
The most instructive precedent is the long and contested debate over DNA
evidence in Islamic family law.  Classical Islamic law establishes paternity
(\textit{nasab}) through the marital bed (\textit{far\={a}sh}), voluntary
acknowledgment, and in contested cases \textit{bayyina} (witness
testimony).  The Qur\textsuperscript{'}anic procedure of \textit{li\textsuperscript{c}\={a}n}
(mutual oath-swearing) provides a mechanism for disputed paternity that
deliberately avoids biological certainty.  DNA testing, which can establish
biological paternity with near-certainty, raised the question of whether it
could constitute \textit{bayyina} or override \textit{li\textsuperscript{c}\={a}n}.
 
The Islamic Fiqh Council of the Muslim World League (Mecca, 2002) issued a
resolution acknowledging DNA testing as an effective scientific method yielding
``certain or near-certain results,'' but specified that it should support
\textit{shar\={\i}\textsuperscript{c}a}-based methods rather than replace them
\citep{islamiclaw_blog2021}.  The Malaysian National Fatwa Council (2012)
similarly restricted DNA's role, favouring traditional methods.  Several
contemporary \textit{fuqah\={a}'} have argued, following Ibn Qayyim
al-Jawziyya's expansive reading of \textit{bayyina} as ``anything that reveals
the truth,'' that DNA should be admitted as an independent evidentiary basis
\citep{ncbi_bioethics2015}.  Others insist that DNA can function only as
\textit{qar\={\i}na} (circumstantial evidence) rather than \textit{bayyina}
proper, and cannot override established presumptions.
 
This debate is directly analogous to the AI consciousness monitoring question: A new scientific method produces
near-certain biological information that disrupts an established legal
framework designed around a different evidentiary basis.  The DNA case shows
that Islamic jurisprudence is capable of accommodating new evidence
types but only after a sustained, contentious, multi-decade deliberative
process, and with significant residual disagreement.  It also shows the
characteristic Islamic response: admit the new evidence as supplementary, not
constitutive. Preserve the existing framework's authority over ultimate
determinations and impose procedural conditions (government-authorised labs,
multiple independent analyses, restrictions on who may request testing) that
manage the technology's disruptive potential \citep{ajis_dna2021}.  A
doctrine of AI-based \textit{bayyina} will need to develop analogous
constraints if it is to be accepted.
 
\textbf{EEG as confirmatory test: a partial precedent within brain death.}
There is a smaller-scale precedent internal to the brain death debate itself.
Saudi Arabia's High Committee on Brain Death already insists on EEG as a
confirmatory test before establishing the diagnosis \citep{albar2012}.  This
represents an implicit jurisprudential judgment: the standard clinical
examination alone does not produce \textit{bayyina} sufficient for death
certification in an Islamic context; an additional neurophysiological test is
required.  This is noteworthy precisely because it shows that Islamic
scholarly bodies have already, in effect, elevated the evidentiary threshold
for brain death beyond what secular protocols require, not on clinical
grounds but on epistemological ones.  AI continuous monitoring can be
understood as the logical extension of this precedent: if a flat-line EEG is
required to supplement clinical signs, what is the status of an AI system
that detects transient neural activity the EEG would miss?  The Saudi
requirement, far from resolving the AI question, sharpens it: the same
precautionary logic that motivated the EEG requirement now generate a demand
for AI monitoring that the EEG precedent cannot itself satisfy.
 
\section{Structural Problems in AI Based Death Detection }
\label{sec:mismatches}
 
\subsection{The Probabilistic Output Problem}
In section 2 we addressed disorders of consciousness for the patients who are not brain-dead. Brain death, by contrast, is a distinct clinical and legal determination: the irreversible cessation of all brain function, including the brainstem. While these are not the same population, the jurisprudential challenge this paper addresses arises at their intersection: AI monitoring systems developed for DoC classification are increasingly being evaluated as confirmatory tools in brain death protocols, and probabilistic outputs generated in that context attach to death certification rather than to treatment planning. The example that follows should be read in this light i.e., a residual probability of neural activity flagged during a brain death evaluation, not a DoC classification in a patient already known to be in MCS. Thus consider, a system might output: 87\% probability
of irreversible loss of integrative brain function; 13\% probability of
residual activity consistent with minimally conscious state.  This is
genuinely more informative and may be more accurate.

AI consciousness scores occupy an intermediate evidentiary position, above shakk but below bayyina. This is so because \textit{Yaqin} (certainity) is unattainable for three independent reasons: first, the system produces a probability distribution, not a verdict, and no threshold transformation of a probability into a binary output preserves the certainty the concept requires. Second, the hard problem of consciousness (Section 4.3) means that even a perfect neural measurement would not establish the presence or absence of the \textit{ruh} (soul). Third, current systems are trained predominantly on European and North American cohorts, introducing distributional uncertainty that compounds the first two. Yet the score exceeds mere \textit{shakk} (doubt) in ways that matter legally. It is reproducible across independent evaluations, externally validated against gold-standard behavioural measures. Islamic legal history accommodates intermediate evidentiary categories: the DNA precedent shows that a new evidence type can be admitted as \textit{qarina} (circumstantial corroboration) without achieving the status of bayyina proper, and without being dismissed as legally ineffective conjecture. AI consciousness scores may be candidates for an analogous category, subject to the procedural conditions discussed in Section 5.

The 13\% residual probability needs to be thought through within an Islamic
precautionary framework.  The \textit{i\d{h}tiy\={a}\d{t}} principle demands
that doubt be resolved in favour of the more protective conclusion.  The AI
makes uncertainty \emph{explicit and quantified}, which paradoxically may
make precautionary withdrawal \emph{less}, not more, permissible under
\textit{fiqh}.  Epistemic uncertainty quantification, a feature widely
advocated in trustworthy clinical AI precisely because it prevents false
confidence and supports informed human oversight, here
produces the opposite of its intended effect. It does so by making the residual
probability of preserved consciousness legible as a number rather than leaving
it as a vague clinical impression, the system transforms what was previously a
matter of professional judgment (and therefore \textit{i\d{h}tiy\={a}\d{t}}'s
domain) into a documented, actionable quantity that the precautionary framework
\emph{cannot} set aside.  In Islamic legal terms, an uncertain clinical
impression might be navigated through the physician's discretion and the
family's \textit{wal\={\i}} authority; a logged 13\% posterior probability of
awareness, attached to a certified death determination, is a different kind of
object. It may generates an obligation to justify why the precautionary
conclusion was not taken, and for which no existing doctrine of computational
\textit{bayyina} yet supplies that justification.
 
\subsection{The Temporal Granularity Problem}
Death in Islamic jurisprudence is a \emph{moment}, an event with a before
and an after.  The \textit{r\={u}\d{h}} departs; the person is gone.  Legal
consequences are attach to this moment: inheritance distributes, marriage
dissolves, organ retrieval becomes permissible. AI continuous monitoring introduces \emph{temporal trajectories of neural
state}.  A monitoring system might show that what appeared as flat neural
activity at 14:00 was preceded by transient fluctuations at 11:00 and
followed by a possible perturbational response at 17:00.  Death becomes not
a discrete event but a process with uncertain boundaries and
probabilistically detected waypoints.
 
\subsection{The Hard Problem Cannot Be Dissolved}
\label{sec:hardproblem}
 
A central claim of Islamic metaphysics is that the \textit{r\={u}\d{h}} is breathed into the human
being by God. The classical Islamic position on whether it has a biological substrate or not is left ambiguous.  The ML community has implicitly adopted the working assumption of neural
correlates of consciousness (NCC) that consciousness supervenes on, or is
identical to, certain neural processes.  This is a \emph{philosophical}
assumption, not an established fact, and it is precisely the assumption that
Islamic theology is agnostic about.  AI consciousness scores cannot
be interpreted as evidence about the \textit{r\={u}\d{h}}'s presence or
absence.
 
\subsection{The \textit{Ijtih\={a}d} Speed Mismatch}
 
Islamic jurisprudence develops through \textit{ijtih\={a}d}, independent
scholarly reasoning, and the issuance of \textit{fatw\={a}s}.  This process
operates on timescales of months to years.  The 2018 FCNA
\textit{fatw\={a}} on organ donation resulted from two years of
multidisciplinary deliberation \citep{fcna2020}. AI clinical capabilities are advancing on timescales of weeks to months.
The PerBrain Consortium's multimodal results appeared as a preprint in
November 2024; the 680,000-sample deep-learning consciousness-detection
system in October 2024.  Each advance potentially shifts the jurisprudential ground under existing \textit{fatw\={a}s}.
There is currently no institutional mechanism in any Islamic scholarly body
for ongoing iterative engagement with rapidly evolving AI capabilities.  The
AI ethics community has begun to discuss ``pacing'' as a governance challenge
\citep{dafoe2018}, but exclusively within secular frameworks.
 
\subsection{The AI Surrogate Problem}
A related trajectory compounds the issues above.  AI-based \emph{patient
preference predictors} (PPPs) i.e., systems that fine-tune large language models
on a patient's prior decisions, values, and behavioural data to infer what
they would have wanted when incapacitated, are now technically feasible and
actively debated \citep{earp2024p4,ppp_nejm2025}.  Proponents argue they
outperform human surrogates, whose accuracy at identifying p  atient
preferences is near chance \citep{ppp_scoping2025}.  Critics counter that
clinical teams may treat probabilistic outputs as determinative in a domain
that is ``relational, existential, and culturally diverse''
\citep{ahmad2025fairness,ppp_ethics2024}.
 
PPPs are designed within the
Western bioethical tradition of \emph{individual autonomy}: their goal is to
extend, computationally, the self-determination of an incapacitated patient.
Islamic end-of-life authority is not located in the patient's preferences but
in a tripartite structure of obligations: the \textit{wal\={\i}} (guardian),
whose duty is to protect the patient's \textit{ma\d{s}la\d{h}a} within
Islamic law rather than to substitute the patient's judgment; the
\textit{tab\={\i}b} (physician), whose \textit{am\={a}na} (trusteeship)
is independent of family or patient preferences; and the \textit{faq\={\i}h}
(jurist), whose consultation is required in complex cases
\citep{eol_islamic_scoping2025,bmc_third_party2025}.  The \textit{wal\={\i}}
asks not ``what would the patient have wanted?'' but ``what does Islamic law
require, and what serves this patient's \textit{ma\d{s}la\d{h}a}?''  These
are different questions that no preference-trained model can answer.  The
accountability gap further matters: Islamic ethics requires every morally
significant act to be traceable to a responsible agent
(\textit{muk\={a}llaf}) who bears \textit{takl\={\i}f} before God. This is a chain
that is obscured, not replaced, by algorithmic recommendations
\citep{ahmad2025fairness}.
 
\section{Implications and Research Agenda}
\label{sec:agenda}
 
\subsection{For Machine Learning Researchers}
\textbf{Transparency about population-level epistemics.}
DoC papers should report population-level uncertainty distributions alongside
aggregate performance metrics, particularly for patients near diagnostic
boundaries. \textbf{Cross-cultural validation.}
Large-scale training datasets are predominantly drawn from European and
North American clinical populations \citep{perbrain2024}.  Muslim-majority
populations are systematically underrepresented in the training data of
systems that will be deployed in their healthcare systems, a specific
instance of the distribution-shift and representational fairness problems
documented in medical AI \citep{ahmad2025fairness}. \textbf{Interpretability in high-stakes settings.}
Deep learning approaches offer ``superior predictive power but with higher
cost of interpretability'' \citep{frontiers_doc2025}.  A decision that
cannot be explained cannot be adjudicated within Islamic legal process.
Investment in interpretable architectures is therefore a jurisprudential as
well as a technical priority. \textbf{Acknowledgement of the hard problem.}
ML consciousness-detection papers should explicitly state that their systems
measure neural correlates of consciousness, not consciousness itself, and
that the relationship between these is a philosophical working assumption,
not an established fact.
 
\subsection{For Islamic Bioethicists}
 
\textbf{Direct engagement with the technical literature.}
Most Islamic bioethics papers on brain death cite clinical criteria that has been superceded \citep{champasha2017}.  To the best of our knowledge, the CMD literature, the PerBrain
trial results, and the deep-learning systems described above have not entered
Islamic bioethics discourse. \textbf{A doctrine of computational \textit{bayyina}.}
A developed doctrine governing probabilistic machine outputs, with explicit
conditions for sufficiency, weighting, and mandatory precautionary
adjustments, may be needed to address problems like the one discussed here. \textbf{Institutional mechanisms for rapid \textit{ijtih\={a}d}.}
Existing Islamic scholarly bodies should establish standing technical
advisory committees with ongoing access to ML researchers and clinical
neurologists.
 
\section{Broader Significance}
\label{sec:broader}
The ML community's engagement with value-aligned and culturally situated
ethics \citep{chuang2022,khan2024} has remained predominantly within
Western philosophical frameworks: principlism, contractarianism, utilitarian
welfare metrics.  The Islamic ethical tradition represents not merely a
different set of cultural values to be accommodated, but a different
epistemological framework that challenges some working assumptions of the ML
ethics enterprise itself. The Qur\textsuperscript{'}an's instruction that the \textit{r\={u}\d{h}} is
beyond human determination is a philosophically serious position about the
underdetermination of soul by substrate and the category error involved in
treating ultimate questions about personhood as tractable by measurement.
The hard problem of consciousness which neuroscience cannot solve
\citep{chalmers1995} represents, in secular philosophical terms, the same
epistemic gap that Islamic theology names with the concept of \textit{ghayb}
(the unseen).
\section{Conclusion}
\label{sec:conclusion}
AI-powered disorders-of-consciousness detection is  a clinical diagnostic advance.  In the context of brain death
determination, it is a jurisprudential event as well. It complicates existing
Islamic legal rulings, challenges traditional Islamic epistemic standards for death
evidence. The AI community has a  responsibility here that is not covered by
standard algorithmic fairness analysis.  Researchers must engage the
jurisprudential literature directly and acknowledge that their systems are
being deployed into a world where the most consequential interpretations of
their outputs will be made within frameworks their models cannot address.

\bibliographystyle{icml2025}
\bibliography{references}


\end{document}